\documentclass[twocolumn,english,twocolumn,english,aps,prb,floats,showpacs]{revtex4}
\usepackage[T1]{fontenc}
\usepackage[latin9]{inputenc}
\usepackage{amssymb}
\usepackage{esint}
\usepackage{graphicx}
\usepackage{bm}
\makeatletter
\@ifundefined{textcolor}{}
{%
 \definecolor{BLACK}{gray}{0}
 \definecolor{WHITE}{gray}{1}
 \definecolor{RED}{rgb}{1,0,0}
 \definecolor{GREEN}{rgb}{0,1,0}
 \definecolor{BLUE}{rgb}{0,0,1}
 \definecolor{CYAN}{cmyk}{1,0,0,0}
 \definecolor{MAGENTA}{cmyk}{0,1,0,0}
 \definecolor{YELLOW}{cmyk}{0,0,1,0}
}

\makeatother

\usepackage{babel}
\begin{document}

\title{Angular momentum in spin-phonon processes}
\author{D. A. Garanin and E. M. Chudnovsky}
\affiliation{Physics Department, Lehman College,
The City University of New York, 250 Bedford Park Boulevard West,
Bronx, NY 10468-1589, U.S.A.}
\date{\today}

\begin{abstract}
Quantum theory of spin relaxation in the elastic environment is revised with account of the concept of a phonon spin recently introduced by Zhang and Niu (PRL 2014). Similar to the case of the electromagnetic field, the division of the angular momentum associated with elastic deformations into the orbital part and the part due to phonon spins proves to be useful for the analysis of the balance of the angular momentum. Such analysis sheds important light on microscopic processes leading to the Einstein - de Haas effect.
\end{abstract}

\maketitle

\section{Introduction}
The problem of conservation of angular momentum in systems containing magnetic moments has been around since the discovery of Einstein - de Haas \cite{EdH} and Barnett \cite{Barnett} effects one hundred years ago. The first effect demonstrated that the change in the magnetic moment  of a feely suspended body generates mechanical rotation, while the second demonstrated that mechanical rotation induces magnetization. For some time the Einstein - de Haas and Barnett effects were used to measure the gyromagnetic ratio of solids \cite{Bar-Scott}. The significance of such measurements was diminished by the discovery of the electron spin resonance and the ferromagnetic resonance that provided a more accurate determination of the gyromagnetic ratio. After that the experiments on macroscopic magneto-mechanical gyroscopic effects have been largely abandoned. Surprisingly, however,  microscopic mechanisms of the transfer of the spin angular momentum to the phonon system and subsequently to the body as a whole remain poorly understood. 

The tradition that goes back to the pioneering work on spin-phonon relaxation by Van Vleck \cite{VanVleck-PR40} consists of ignoring conservation of angular momentum under the excuse that the Hamiltonian of the system does not possess full rotational invariance. It is clear, however, that in theory (and in experiment) the angular momentum in a system of interacting spins and phonons is conserved. This prompted a significant effort by a number of researchers to formulate the theory of magneto-elastic interactions in a rotationally invariant manner \cite{mel72prl,dohful75,fed75prb,bonmel76prb,fedmel76prb,fed77prb,chugarsch05prb}. The advantage of such approach is that it is parameter free in a sense that spin-phonon rates can be expressed in terms of the well-known independently measured parameters. 

Emergence of micro- and nanoelectromechanical devices (MEMS and NEMS) rejuvinated interest to the problem of angular momentum in magneto-mechanical systems \cite{chugar14prb}. Einstein - de Haas effect at the nanoscale has been experimentally studied in magnetic microcantilevers \cite{walmorkab06apl,limimtwal14epl} and theoretically explained by the motion of domain walls \cite{jaachugar09prb}. Switching of magnetic moments by mechanical torques in nanocantilevers has been proposed \cite{Kovalev-PRL2005,cai-PRA14,CJ-JAP2015}.  Mechanical resonators containing single magnetic molecules have been studied by quantum methods \cite{jaachu09prl,jaachugar10epl,kovhaybau11prl,carchu11prx,okechugar13prb}.  Experiments have progressed to the measurement of the angular momentum exchange between of a single molecular spin and a carbon nanotube \cite{ganklyrub13NatNano,ganklyrub13acsNano}. 

In nanoresonators the problem is somewhat simpler due to the finite number of resonant modes. For a single spin in a macroscopic body, however, the number of phonon degrees of freedom is practically infinite. In relation to the angular momentum this problem has received significant recent attention in experiments with atomic spin - based qubits \cite{Awschalom1,Awschalom2} and in application to spintronics \cite{Qin-PRB12}. To address this problem Zhang and Niu recently introduced a concept of the phonon spin \cite{Zhang-Niu}. 

\begin{figure}[htbp!]
\center
\includegraphics[scale=0.35]{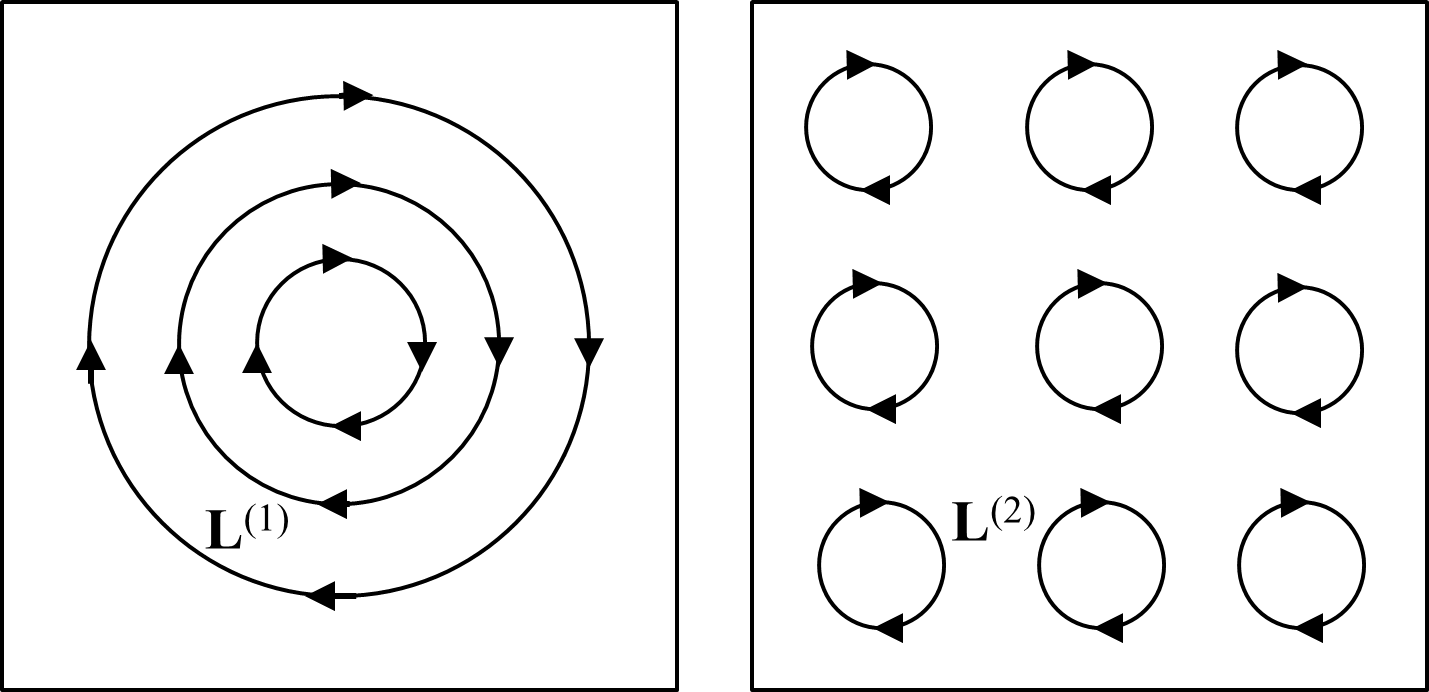}
\caption{Conceptual representation of the motion of the elastic medium that generates the orbital angular momentum ${\bf L}^{(1)}$ and the phonon spin angular momentum ${\bf L}^{(2)}$.}
\label{Fig1}
\end{figure}

In this paper we investigate this concept for the process of the relaxation of a single atomic spin in a macroscopic body. By developing an approach similar to that for photons we find that within the elastic theory the angular momentum of phonons can be naturally split into the orbital angular momentum ${\bf L}^{(1)}$ and the spin angular momentum ${\bf L}^{(2)}$. The orbital part corresponds to the rotation of the elastic medium around a certain point, while the spin part corresponds to a small-radius circular shear displacements of points of the elastic media around their equilibrium positions, see Fig. \ref{Fig1}. 

The paper is structured as follows. The concept of the angular momentum in classical and quantum theories of elasticity is discussed in Section \ref{AM}. Conservation of the total angular momentum is studied in Section \ref{conservation} by computing its commutator with the Hamiltonian. Quantum dynamics of the angular momentum of the relaxing spin and emitted phonons is investigated in Section \ref{QT}. Section \ref{conclusion} contains summary of the results and some final comments.

\section{The angular momentum}\label{AM}

\subsection{Angular momentum in the classical theory of elasticity}
The angular momentum of the elastic solid is defined as
\begin{equation}\label{L}
\mathbf{L}=\int d^{3}r\left(\mathbf{r}+\mathbf{u}\right)\times\mathbf{p},
\end{equation}
where time-indendent $\mathbf{r}$ corresponds to the non-deformed
body, $\mathbf{u}({\bf r},t)$ is deformation, and $\mathbf{p}({(\bf r},t) = \rho \dot{\bf u}({(\bf r},t)$ is the momentum
density. It consists of two parts
\begin{equation}
{\bf L} = {\bf L}^{(1)} + {\bf L}^{(2)} ,
\end{equation}
where
\begin{equation}
{\bf L}^{(1)} = \int d^{3}r \rho \,\mathbf{r}\times \dot{\bf u}, \quad
{\bf L}^{(2)} = \int d^{3}r \rho\,\mathbf{u}\times \dot{\bf u}.
\end{equation}
The orbital part described by ${\bf L}^{(1)}$ corresponds to the rotation of the elastic medium around the origin, while the spin part described by ${\bf L}^{(2)}$ corresponds to a small-radius circular shear displacements of points of the elastic media around their equilibrium positions, see Fig. \ref{Fig1}.

Applying time derivative to these expressions one obtains
\begin{equation}\label{Lz-derivative}
\dot{\bf L}^{(1)} = \int d^3 r \rho \, {\bf r}  \times \ddot {\bf u}, \quad \dot{\bf L}^{(2)} = \int d^3 r \rho \,{\bf u}  \times \ddot {\bf u}.
\end{equation}
The dynamical equation for the displacement field is the Newton's equation 
\begin{equation}\label{elastic}
\rho \frac{\partial^2 u_{\alpha}}{\partial t^2} = \frac{\partial
\sigma_{\alpha \beta}}{\partial r_{\beta}} 
\end{equation}
with the force in the right-hand-side being a gradient of the stress tensor 
$\sigma_{\alpha \beta} = {\delta H}/{\delta e_{\alpha\beta}}$. Here 
$H$ is the Hamiltonian of the system and $e_{\alpha \beta} ={\partial u_{\alpha}}/{\partial r_{\beta}}$.
After integrating by parts in equations (\ref{Lz-derivative}) and assuming zero elastic stress at the boundary of the body, one obtains
\begin{equation}\label{dotL-sigma}
\dot{L}_{\alpha}^{(1)} = -\int d^3 r \epsilon_{\alpha \beta \gamma}\sigma_{\gamma\beta} , \quad \dot{L}_{\alpha}^{(2)} = -\int d^3 r\epsilon_{\alpha \beta \gamma} e_{\beta\delta}\sigma_{\gamma\delta}
\end{equation} 
In the linear elastic theory $\dot{\bf L}^{(1)}$ is zero in the absence of internal torques due to the symmetry of the stress tensor. Such torques are ignored by the conventional elastic theory. ${\bf L}^{(2)}$, that is quadratic on deformations, is also neglected in the linear elastic theory, making $\dot{\bf L} = 0$. 

When $\sigma_{\alpha \beta}$ is non-symmetric, more care is needed. This can happen for two reasons. The first reason is the intrinsic anharmonicity of the elastic theory due to the nonlinearity of the strain tensor \cite{LL-elasticity},
 \begin{equation}\label{strain-nonlinear}
u_{\rho\eta} = \frac{1}{2}\left(e_{\rho\eta} + e_{\eta\rho} + e_{\nu\rho}e_{\nu\eta}\right)
\end{equation}
The fact that $H$ must depend on  $u_{\rho\eta}$ leads to 
\begin{equation}
\sigma_{\gamma\delta} = \frac{\delta H}{\delta e_{\gamma\delta}} = \frac{\delta u_{\rho\eta}}{\delta e_{\gamma\delta}} \frac{\delta {H}}{\delta u_{\rho\eta}} = \frac{\delta {H}}{\delta u_{\gamma\delta}} + e_{\gamma\rho}\frac{\delta {H}}{\delta u_{\rho\delta}}
\end{equation}
which is non-symmetric. It is easy to see that in this case the second term in Eq.\ (\ref{dotL-sigma}) is needed for the condition $\dot{\bf L} = 0$ to be exact. 

Anharmonicity, however, is not the only reason for $\sigma_{\alpha \beta}$ to be non-symmetric. It also happens in the presence of spins because spin dynamics generates internal torques. Consider, e.g., a uniaxial spin Hamiltonian of the form
\begin{equation}
\hat{H}=-D\left(\mathbf{n\cdot S}\right)^{2}
\end{equation}
with ${\bf n}$ being the magnetic anisotropy axis. Elastic deformations of the body rotate
the anisotropy axis $\mathbf{n}$ by a small angle ${\bm \phi}$ 
\begin{equation}
{\bm \phi}=\frac{1}{2}\nabla\times\mathbf{u}(\mathbf{r}), \quad \phi_\alpha = \frac{1}{2}\epsilon_{\alpha\beta\gamma}e_{\gamma\beta}.\label{phi_def}
\end{equation}
To the first order in ${\bm \phi}$ one has $\mathbf{n}=\mathbf{e}_{z}+\left[{\bm \phi}\times\mathbf{e}_{z}\right]$.
Expanding $\hat{H}$ up to the linear terms in $\phi$, we get
$\hat{H}=\hat{H}_{A}+\hat{H}_{\mathrm{s-ph}}$, where $\hat{H}_{A}=-DS_{z}^{2}$ and the spin-lattice coupling
is given by \cite{chugar97prb}
\begin{equation}\label{H1}
\hat{H}_{\mathrm{s-ph}}=-D\left(S_{x}S_{z}+S_{z}S_{x}\right)\phi_{y}+D\left(S_{y}S_{z}+S_{z}S_{y}\right)\phi_{x}.
\end{equation}
The corresponding stress tensor $\sigma_{\alpha \beta} = {\delta H}_{\mathrm{s-ph}}/{\delta e_{\alpha\beta}}$ is non-symmetric. Writing it as 
\begin{equation}
\sigma_{\alpha \beta} = \frac{\delta H_{\mathrm{s-ph}}}{\delta e_{\alpha\beta}} = \frac{\delta H_{\mathrm{s-ph}}}{\delta \phi_\gamma} \frac{\delta \phi_\gamma}{\delta e_{\alpha\beta}}
\end{equation}
one obtains
\begin{equation}
\dot{\bf L}^{(1)} = -\int d^3 r \frac{\delta H_{\mathrm{s-ph}}}{\delta {\bm \phi}},
\end{equation}
which explicitly expresses the internal mechanical torque via the elastic twist. The latter comes from the spin-lattice coupling. In what follows we will show that ${\bf L}^{(2)}$ associated with the phonon spin is also generated in the problem of quantum relaxation of the atomic spin. Consequently, ${\bf L}^{(2)}$, that is usually neglected in the linear elastic theory, turns out to be important for the conservation of the total angular momentum, even in cases when the problem is  solved with linear non-interacting phonons.

\subsection{Quantum theory of phonon angular momentum}
To obtain the second-quantized expression for the angular
momentum, we use canonical quantization of phonons
\begin{equation}
\mathbf{u}(\mathbf{r})=\sqrt{\frac{\hbar}{2\rho V}}\sum_{\mathbf{k}\lambda}\frac{\mathbf{e}_{\mathbf{k}\lambda}e^{i\mathbf{k\cdot r}}}{\sqrt{\omega_{\mathbf{k}\lambda}}}a_{\mathbf{k}\lambda}+h.c.,\label{uQuantized}
\end{equation}
where $\rho$ is the mass density, $V$ is the volume, $\mathbf{e}_{\mathbf{k}\lambda}$ are
polarization vectors, $\omega_{\mathbf{k}\lambda}$ are phonon frequencies
and $a$, $a^{\dagger}$ are creation and annihilation operators of
phonons. One uses Eq. (\ref{uQuantized}), as well as 
\begin{equation}
\mathbf{p}(\mathbf{r})=\rho\mathbf{\dot{u}}(\mathbf{r})=-i\sqrt{\frac{\rho\hbar}{2V}}\sum_{\mathbf{k}\lambda}\mathbf{e}_{\mathbf{k}\lambda}\sqrt{\omega_{\mathbf{k}\lambda}}e^{i\mathbf{k\cdot r}}a_{\mathbf{k}\lambda}+h.c.\label{p_quantized}
\end{equation}

The angular momentum of the body, Eq.\ (\ref{L}), consists of two contributions,
$\hat{\mathbf{L}}=\mathbf{\hat{L}}^{(1)}+\mathbf{\hat{L}}^{(2)}$ that have been discussed earlier.
Here $\mathbf{\hat{L}}^{(1)}$ is first order in phonon operators and it can be interpreted as the orbital angular momentum of the phonons. The term $\mathbf{\hat{L}}^{(2)}$ is second-order in phonon operators and it can be interpreted as the spin of the phonons.
Splitting the angular momentum into two parts is similar
to that of photons. It will be shown below that the spin of a phonon
is $\hbar$ and the phonon-spin eigenstates are circularly-polarized
phonons.

The operator of the orbital angular momentum becomes
\begin{equation}
\mathbf{\hat{L}}^{(1)}=\sqrt{\frac{\rho\hbar}{2V}}\sum_{\mathbf{k}\lambda}\sqrt{\omega_{\mathbf{k}\lambda}}\left[\mathbf{e}_{\mathbf{k}\lambda}\times\mathbf{j}_{\mathbf{k}}\right]a_{\mathbf{k}\lambda}+h.c.,\label{L1_via_jk}
\end{equation}
where $\mathbf{j}_{\mathbf{k}}\equiv i\int d^{3}r\,\mathbf{r}e^{i\mathbf{k\cdot r}}$.
As, by symmetry, $\mathbf{j}_{\mathbf{k}}$ can only be directed along
$\mathbf{k}$, only transverse phonons contribute into $\mathbf{\hat{L}}^{(1)}$.
In an infinite body, wave vectors are continuous, so that one can
replace summation by integration
\begin{equation}
\frac{1}{V}\sum_{\mathbf{k}}\ldots\Rightarrow\int\frac{d^{3}k}{\left(2\pi\right)^{3}}\ldots\label{summation_to_integration}
\end{equation}
Then one can express $\mathbf{j}_{\mathbf{k}}$ as
\begin{equation}
\mathbf{j}_{\mathbf{k}}=\left(2\pi\right)^{3}\partial_{\mathbf{k}}\delta\left(\mathbf{k}\right).\label{j_k_via_delta_k}
\end{equation}

Dropping the terms $aa$ and $a^{\dagger}a^{\dagger}$ in $\mathbf{\hat{L}}^{(2)}$ that do not conserve
the number of phonon excitations, one obtains after integration over the volume 
\begin{equation}
\mathbf{\hat{L}}^{(2)}=\frac{i\hbar}{2}\sum_{\mathbf{k}\lambda\lambda'}\left[\mathbf{e}_{\mathbf{k}\lambda}\times\mathbf{e}_{\mathbf{k}\lambda'}\right]a_{\mathbf{k}\lambda}a_{\mathbf{k}'\lambda'}^{\dagger}+h.c.
\end{equation}
Keeping only transverse phonons, $\lambda\lambda'=1,2$ and using
$\left[\mathbf{e}_{\mathbf{k}1}\times\mathbf{e}_{\mathbf{k}2}\right]=\mathbf{k}/k$,
one arrives at
\begin{equation}
\mathbf{\hat{L}}^{(2)}=i\hbar\sum_{\mathbf{k}}\frac{\mathbf{k}}{k}\left(a_{\mathbf{k}2}^{\dagger}a_{\mathbf{k}1}-a_{\mathbf{k}1}^{\dagger}a_{\mathbf{k}2}\right).\label{L2_via_12}
\end{equation}
This operator becomes diagonal in terms of numbers of circularly-polarized
phonons $a_{\mathbf{k}\pm}\equiv\left(a_{\mathbf{k}1}\pm ia_{\mathbf{k}2}\right)/\sqrt{2}$
\begin{equation}
\mathbf{\hat{L}}^{(2)}=\hbar\sum_{\mathbf{k}}\frac{\mathbf{k}}{k}\left(-a_{\mathbf{k}+}^{\dagger}a_{\mathbf{k}+}+a_{\mathbf{k}-}^{\dagger}a_{\mathbf{k}-}\right).\label{L2_via_a_pm}
\end{equation}
Each such phonon carries an angular momentum $\hbar$ parallel or
anti-parallel to its wave vector that can be interpreted as the spin
of the phonon.

In what follows we will study conservation of the angular momentum in the spin-relaxation model by computing its commutator with the Hamiltonian. Introducing spin operators $S_{\pm}\equiv S_{x}\pm iS_{y}$ that follow commutation relations $\left[S_{\pm},S_{z}\right]=\pm S_{\pm}$, one obtains from Eq.\ (\ref{H1})
\begin{equation}
\hat{H}_{\mathrm{s-ph}}=-\frac{iD}{2}\left(S_{+}S_{z}+S_{z}S_{+}\right)\phi_{-}+h.c.,\label{H_s-ph}
\end{equation}
where $\phi_{\pm}\equiv\phi_{x}\pm i\phi_{y}$. Using Eqs. (\ref{phi_def})
and (\ref{uQuantized}) with the atomic spin located at $\mathbf{r}=0$, one
obtains
\begin{equation}
\phi_{\pm}=\frac{1}{2}\sqrt{\frac{\hbar}{2\rho V}}\sum_{\mathbf{k}\lambda}\frac{\mathbf{e}_{\pm}\cdot\left[i\mathbf{k}\times\mathbf{e}_{\mathbf{k}\lambda}\right]}{\sqrt{\omega_{\mathbf{k}\lambda}}}\left(a_{\mathbf{k}\lambda}-a_{\mathbf{k}\lambda}^{\dagger}\right),\label{phi_pm}
\end{equation}
where $\mathbf{e}_{\pm}\equiv\mathbf{e}_{x}\pm i\mathbf{e}_{y}$.

\section{Conservation of the angular momentum}\label{conservation}

Let us now check conservation of the total angular momentum
\begin{equation}
\mathbf{J}=\mathbf{L}+\hbar\mathbf{S}
\end{equation}
that implies that $\mathbf{J}$ must commute with the Hamiltonian.
The dynamical change of the spin operator has to be absorbed by the
angular momentum of the elastic matrix, whose evolution is given by
\begin{equation}
\hat{\dot{\mathbf{L}}}=\frac{i}{\hbar}\left[\hat{H}_{\mathrm{s-ph}},\mathbf{\hat{L}}\right].\label{L_dot_comm}
\end{equation}
In particular, the precession of the spin around the anisotropy axis creates the
co-wiggling of the elastic matrix with the spin.

It turns out that by commuting operators one can prove conservation
of some parts of the angular momentum, whereas the complete prove
of conservation requires a full quantum-mechanical solution for the
relaxing spin and phonons created by its precession, presented in
the next section. The situation is different for the angular momentum
components perpendicular and parallel to the anisotropy axis. We will
need commutators
\begin{equation}
\left[\phi_{\pm},\mathbf{\hat{L}}^{(1)}\right]=i\frac{\hbar}{2V}\sum_{\mathbf{k}\lambda}\left(\mathbf{e}_{\pm}\cdot\left[\mathbf{k}\times\mathbf{e}_{\mathbf{k}\lambda}\right]\right)\left[\mathbf{e}_{\mathbf{k}\lambda}\times\mathbf{j}_{\mathbf{k}}\right]\label{phi_L1_comm}
\end{equation}
and
\begin{equation}
\left[\phi_{\pm},\mathbf{\hat{L}}^{(2)}\right]=\frac{\hbar}{2}\sqrt{\frac{\hbar}{2\rho V}}\sum_{\mathbf{k}\lambda}\frac{\mathbf{k}}{\sqrt{\omega_{\mathbf{k}}}}\left(\mathbf{e}_{\pm}\cdot\mathbf{e}_{\mathbf{k}\lambda}\right)\left(a_{\mathbf{k}\lambda}-a_{\mathbf{k}\lambda}^{\dagger}\right)\label{phi_L2_comm}
\end{equation}
that follow from Eqs. (\ref{phi_pm}), (\ref{L1_via_jk}), and (\ref{L2_via_12}).

Let us first consider dynamics of the transverse components of the
angular momentum. The dominant source of spin precession around the
anisotropy axis is the unperturbed spin Hamiltonian $\hat{H}_{A}$:
\begin{equation}
\dot{S}_{x}=\frac{i}{\hbar}\left[\hat{H}_{\mathrm{A}},S_{x}\right]=-\frac{i}{\hbar}D\left[S_{z}^{2},S_{x}\right]=\frac{D}{\hbar}\left(S_{z}S_{y}+S_{y}S_{z}\right).\label{dot_Sx}
\end{equation}
For the matrix, let us first consider the dynamics of the phonon orbital
angular momentum $\hat{\mathbf{L}}^{(1)}$. From Eq. (\ref{phi_L1_comm})
with the help of the identity
\begin{equation}
\sum_{\lambda=1,2}\left(\mathbf{e}_{\mathbf{k}\lambda}\cdot\mathbf{A}\right)\left(\mathbf{e}_{\mathbf{k}\lambda}\cdot\mathbf{B}\right)=\mathbf{A\cdot B}-\left(\frac{\mathbf{k}}{k}\cdot\mathbf{A}\right)\left(\frac{\mathbf{k}}{k}\cdot\mathbf{B}\right)
\end{equation}
and Eq. (\ref{j_k_via_delta_k}) one obtains 
\begin{eqnarray}
\left[\phi_{\pm},\hat{L}_{x}^{(1)}\right] & = & i\frac{\hbar}{2V}\sum_{\mathbf{k}\lambda}\left(\mathbf{e}_{\pm}\cdot\left[\mathbf{k}\times\mathbf{e}_{\mathbf{k}\lambda}\right]\right)\left(\mathbf{e}_{x}\cdot\left[\mathbf{e}_{\mathbf{k}\lambda}\times\mathbf{j}_{\mathbf{k}}\right]\right)\nonumber \\
 & = & i\frac{\hbar}{2V}\sum_{\mathbf{k}\lambda}\left(\mathbf{e}_{\mathbf{k}\lambda}\cdot\left[\mathbf{e}_{\pm}\times\mathbf{k}\right]\right)\left(\mathbf{e}_{\mathbf{k}\lambda}\cdot\left[\mathbf{j}_{\mathbf{k}}\times\mathbf{e}_{x}\right]\right)\nonumber \\
 & = & i\frac{\hbar}{2V}\sum_{\mathbf{k}}\left[\mathbf{e}_{\pm}\times\mathbf{k}\right]\cdot\left[\mathbf{j}_{\mathbf{k}}\times\mathbf{e}_{x}\right]\nonumber \\
 & = & i\frac{\hbar}{2V}\sum_{\mathbf{k}}\left\{ \left(\mathbf{e}_{\pm}\cdot\mathbf{j}_{\mathbf{k}}\right)\left(\mathbf{e}_{x}\cdot\mathbf{k}\right)-\left(\mathbf{e}_{\pm}\cdot\mathbf{e}_{x}\right)\left(\mathbf{k}\cdot\mathbf{j}_{\mathbf{k}}\right)\right\} \nonumber \\
 & = & i\frac{\hbar}{2}\int d^{3}k\left\{ k_{x}\partial_{k_{x}}\delta\left(\mathbf{k}\right)-\left(\mathbf{k}\cdot\partial_{\mathbf{k}}\delta\left(\mathbf{k}\right)\right)\right\} \nonumber \\
 & = & i\hbar.\label{L1_phi_comm}
\end{eqnarray}
Now from Eqs. (\ref{L_dot_comm}) and (\ref{H_s-ph}) one obtains
\begin{equation}
\hat{\dot{L}}_{x}^{(1)}=-D\left(S_{y}S_{z}+S_{z}S_{y}\right).
\end{equation}
Combining this with Eq. (\ref{dot_Sx}), one obtains the conservation
law
\begin{equation}
\hbar\dot{S}_{x}+\hat{\dot{L}}_{x}^{(1)}=0.\label{Conservation_transverse}
\end{equation}
In the same way one can obtain $\hbar\dot{S}_{y}+\hat{\dot{L}}_{y}^{(1)}=0.$

However, Eq. (\ref{Conservation_transverse}) is not the whole story.
One has to consider $\hat{\dot{L}}_{x,y}^{(2)}$ using Eqs. (\ref{phi_L2_comm})
and (\ref{H_s-ph}). The resulting expression is a sum over $\mathbf{k}$,
linear in phonon operators. It is of the same order as the contribution
to $\hbar\dot{S}_{x,y}$ due to the spin-phonon interaction, $i\left[\hat{H}_{\mathrm{s-ph}},S_{x,y}\right]$,
that was ignored above. Both terms discussed here are much smaller
than the dominant terms in the angular momentum, conserved according
to Eq. (\ref{Conservation_transverse}). These small terms are related
to the spin-lattice relaxation of the spin. It is impossible to prove conservation of these terms without performing the full
quantum-mechanical solution of the problem of spin relaxation.

Considering dynamics of the longitudinal component of the angular
momentum, one can prove 
\begin{equation}
\dot{L}_{z}^{(1)}=\frac{i}{\hbar}\left[\hat{H}_{\mathrm{s-ph}},L_{z}^{(1)}\right]=0\label{L1z_dot_zero}
\end{equation}
by a calculation similar to that in Eq. (\ref{L1_phi_comm}). The
terms $\dot{S}_{z}$ and $\dot{L}_{z}^{(2)}$ are related to spin-lattice
relaxation and they are sums over $\mathbf{k}$, linear in phonon
operators. However, one cannot prove

\begin{equation}
\hbar\dot{S}_{z}+\dot{L}_{z}^{(2)}=0
\end{equation}
without the full solution of the quantum problem that will be presented
below.

\section{Quantum theory of the relaxing spin}\label{QT}

\subsection{General solution}

To facilitate solving the problem of spin-lattice relaxation, we reduce
the spin-phonon Hamiltonian to the rotating-wave approximation (RWA)
form that conserves the energy. Consider transitions of the spin $\left|m-1\right\rangle \rightarrow\left|m\right\rangle $
for $m>0$ decreasing its energy and call the spin states $\left|1\right\rangle $
and $\left|0\right\rangle $, respectively. With the help of Eq. (\ref{H_s-ph})
one obtains the spin matrix element of this transition
\begin{equation}
\left\langle m-1\left|\hat{H}_{\mathrm{s-ph}}\right|m\right\rangle =\frac{iD}{2}\left(2m-1\right)l_{m-1,m}\phi_{+},
\end{equation}
where $l_{m-1,m}\equiv\sqrt{S\left(S+1\right)-m(m-1)}$. Using Eqs.
(\ref{H_s-ph}) and (\ref{phi_pm}), one obtains the RWA coupling
in the form
\begin{equation}
\hat{V}=\sum_{\mathbf{k}\lambda}\left(A_{\mathbf{k}\lambda}^{\ast}X^{01}a_{\mathbf{k}\lambda}^{\dagger}+A_{\mathbf{k}\lambda}X^{10}a_{\mathbf{k}\lambda}\right),\label{VRWA}
\end{equation}
where
\begin{equation}
A_{\mathbf{k}\lambda}\equiv-\frac{D}{4}\left(2m-1\right)l_{m-1,m}\sqrt{\frac{\hbar}{2\rho V}}\frac{\mathbf{e}_{+}\cdot\left[\mathbf{k}\times\mathbf{e}_{\mathbf{k}\lambda}\right]}{\sqrt{\omega_{\mathbf{k}\lambda}}}\label{Aklam_def}
\end{equation}
and the $X$-operators are defined by 
\begin{equation}
X^{01}\left|1\right\rangle =\left|0\right\rangle ,\qquad X^{10}\left|0\right\rangle =\left|1\right\rangle .
\end{equation}

The quantum state of the system can be specified by
\begin{equation}
\Psi=\left(cX^{10}+\sum_{\mathbf{k}\lambda}c_{\mathbf{k}\lambda}a_{\mathbf{k}\lambda}^{\dagger}\right)\left|00\right\rangle ,\label{PsiLowExc}
\end{equation}
where $\left|00\right\rangle $ is the ``vacuum'' state. $\Psi$
has only one excitation, spin or phonon. Considering the excited state
of the spin as the reference-energy state, one obtains the Schr\"{o}dinger
equation for the coefficients 
\begin{eqnarray}
\dot{c} & = & -\frac{i}{\hbar}\sum_{\mathbf{k}\lambda}A_{\mathbf{k}\lambda}c_{\mathbf{k}\lambda}\nonumber \\
\dot{c}_{\mathbf{k}\lambda} & = & -i\left(\omega_{\mathbf{k}\lambda}-\omega_{0}\right)c_{\mathbf{k}\lambda}-\frac{i}{\hbar}A_{\mathbf{k}\lambda}^{\ast}c,\label{SELowExc-1}
\end{eqnarray}
where $\omega_{0}\equiv\left(E_{1}-E_{0}\right)/\hbar$ is the frequency of the transition between the spin levels. 

One can integrate the equations for the phonon modes $c_{\mathbf{k}}$
assuming the initial condition of the phonon vacuum: 
\begin{eqnarray}
c_{\mathbf{k}\lambda}(t) & = & -\frac{iA_{\mathbf{k}\lambda}^{\ast}}{\hbar}\int_{0}^{t}dt^{\prime}e^{-i\left(\omega_{\mathbf{k}\lambda}-\omega_{0}\right)(t-t^{\prime})}c(t^{\prime})\nonumber \\
 & = & -\frac{iA_{\mathbf{k}\lambda}^{\ast}}{\hbar}\int_{0}^{t}d\tau e^{-i\left(\omega_{\mathbf{k}\lambda}-\omega_{0}\right)\tau}c(t-\tau)\label{ckIntegrated-1}
\end{eqnarray}
and insert the result into the equation for the spin $c$: 
\begin{equation}
\frac{dc}{dt}=-\frac{1}{\hbar^{2}}\sum_{\mathbf{k}\lambda}\left|A_{\mathbf{k}\lambda}\right|^{2}\int_{0}^{t}d\tau e^{-i\left(\omega_{\mathbf{k}\lambda}-\omega_{0}\right)\tau}c(t-\tau).
\end{equation}
In this integro-differential equation, $c(t-\tau)$ is a slow function
of time, whereas the memory function $f(\tau)=\sum_{\mathbf{k}\lambda}\left|A_{\mathbf{k}\lambda}\right|^{2}e^{-i\left(\omega_{\mathbf{k}\lambda}-\omega_{0}\right)\tau}$
is sharply peaked at $\tau=0.$ Thus one can replace $c(t-\tau)\Rightarrow c(t),$
after which integration over $\tau$ and keeping only real contribution
responsible for the relaxation yields the equation 
\begin{equation}
\frac{dc}{dt}=-\frac{\Gamma}{2}c,\label{cEqRelax}
\end{equation}
and thus 
\begin{equation}
c=e^{-\left(\Gamma/2\right)t},
\end{equation}
where 
\begin{equation}
\Gamma=\frac{2\pi}{\hbar^{2}}\sum_{\mathbf{k}\lambda}\left|A_{\mathbf{k}\lambda}\right|^{2}\delta\left(\omega_{\mathbf{k}\lambda}-\omega_{0}\right)\label{GammaSingSp}
\end{equation}
is the spin relaxation rate. Now, adopting this solution in Eq. (\ref{ckIntegrated-1})
and integrating over time, one obtains for the phonons 
\begin{equation}
c_{\mathbf{k}\lambda}(t)=\frac{A_{\mathbf{k}\lambda}^{\ast}}{\hbar}\frac{e^{-i\left(\omega_{\mathbf{k}\lambda}-\omega_{0}\right)t}-e^{-\left(\Gamma/2\right)t}}{\omega_{\mathbf{k}\lambda}-\omega_{0}+i\Gamma/2}.\label{cklam_res}
\end{equation}

\subsection{Dynamics of the phonon-spin angular momentum}

Let us now compute the phonon-spin angular momentum $L_{z}^{(2)}$
resulting from the relaxation of the spin. Remember that $L_{z}^{(1)}=0$
according to Eq. (\ref{L1z_dot_zero}). It is not neccessary to use
circularly polarized phonons: one can work with linearly polarized
phonons using Eq. (\ref{L2_via_a_pm}) and Eq. (\ref{PsiLowExc}).
For the quantum expectation value one obtains 
\begin{equation}
\mathbf{L}^{(2)}=i\hbar\sum_{\mathbf{k}}\frac{\mathbf{k}}{k}\left(c_{\mathbf{k}2}^{*}c_{\mathbf{k}1}-c_{\mathbf{k}1}^{*}c_{\mathbf{k}2}\right).\label{L2_via_cklam}
\end{equation}
Using Eq. (\ref{cklam_res}) and setting $\omega_{\mathbf{k}\lambda}\Rightarrow\omega_{\mathbf{k}}$,
one obtains
\begin{eqnarray}
\mathbf{L}^{(2)} & = & \frac{i}{\hbar}\sum_{\mathbf{k}}\frac{\mathbf{k}}{k}\left(A_{\mathbf{k}2}A_{\mathbf{k}1}^{\ast}-A_{\mathbf{k}2}A_{\mathbf{k}1}^{\ast}\right)\nonumber \\
 & \times & \frac{1+e^{-\Gamma t}-\left(e^{-i\left(\omega_{\mathbf{k}}-\omega_{0}\right)t}+e^{i\left(\omega_{\mathbf{k}}-\omega_{0}\right)t}\right)e^{-\left(\Gamma/2\right)t}}{\left(\omega_{\mathbf{k}}-\omega_{0}\right)^{2}+\Gamma^{2}/4}. \nonumber \\ \label{L2_t}
\end{eqnarray}
In the integration over $\mathbf{\omega_{\mathbf{k}}}$, one goes
to the upper and lower complex half-plane for the two different
oscillating terms. As the result one obtains 
\begin{equation}
\mathbf{L}^{(2)}=\frac{2\pi}{\hbar\Gamma}\left(1-e^{-\Gamma t}\right)\sum_{\mathbf{k}}\frac{\mathbf{k}}{k}\delta\left(\omega_{\mathbf{k}}-\omega_{0}\right)\left(iA_{\mathbf{k}2}A_{\mathbf{k}1}^{\ast}+h.c.\right).\label{L2_via_A}
\end{equation}
It remains to show that the integral over $\mathbf{k}$ in this expression
can be expressed through $\Gamma$ so that $\Gamma$ cancels and the
result simplifies. Indeed, the combination that enters Eq. (\ref{GammaSingSp})
after simplifications becomes
\begin{equation}
\left|A_{\mathbf{k}1}\right|^{2}+\left|A_{\mathbf{k}2}\right|^{2}=D^{2}\left[\left(2m-1\right)l_{m-1,m}\right]^{2}\frac{\hbar}{4\rho V}\frac{k_{z}^{2}}{\omega_{\mathbf{k}}}.\label{A_combination_Gamma}
\end{equation}
On the other hand, in Eq. (\ref{L2_via_A}) one obtains
\begin{equation}
iA_{\mathbf{k}2}A_{\mathbf{k}1}^{\ast}+h.c.=-D^{2}\left[\left(2m-1\right)l_{m-1,m}\right]^{2}\frac{\hbar}{4\rho V}\frac{kk_{z}}{\omega_{\mathbf{k}}}.
\end{equation}
Note that in Eq. (\ref{L2_via_A}) only the longitudinal component
$L_{z}^{(2)}$ is non-zero by symmetry. The latter is just the negative
of Eq. (\ref{A_combination_Gamma}) that enters $\Gamma$, Eq. (\ref{GammaSingSp}).
Thus in Eq. (\ref{L2_via_A}) $\Gamma$ cancels out and one obtains
the simple behavior 
\begin{equation}
L_{z}=L_{z}^{(2)}=-\left(1-e^{-\Gamma t}\right)\hbar,
\end{equation}
as the spin undergoes a relaxational transition $\left|m-1\right\rangle \rightarrow\left|m\right\rangle $.
This means that the total angular momentum in the system spin + phonons
is conserved. 

\section{Discussion}\label{conclusion}
We have analyzed the transfer of the angular momentum from the atomic spin to the orbital and spin angular momentum of phonons. These two parts of the angular momentum of the phonon system are clearly distinguishable. The orbital part is first order on the phonon operators. Its classical counterpart is the twist of the elastic matrix around the position of the atomic spin, which is linear on the displacement field. The spin part of the phonon angular momentum is second order on phonon operators. Its classical counterpart  corresponds to the rotational shear deformations that are quadratic on the diplacement field.

Conservation of the angular momentum in the process of the relaxation of the atomic spin has been demonstrated by us explicitly. It turns out that the change in the transverse part of the atomic spin is balanced by the orbital part of the phonon angular momentum, while the change in the relaxing longitudinal part of the atomic spin is balanced by the spin part of the phonon angular momentum. These findings can be useful in schemes where individual atomic spins (e.g., used as qubits) are manipulated by phonons. 

An outstanding problem, not addressed in this paper, is how the orbital and spin angular momenta carried by phonons get transferred to the rotation of the body as a whole in the Einstein - de Haas effect. To answer this question one must recall that in a typical Einstein - de Haas experiment one induces rotational oscillations of a macroscopic body by the low frequency ac magnetic field. The corresponding time scales are much greater than lifetimes of phonons emitted in atomic spin transitions. Consequently such phonons fully equilibrate on the time scale of the transfer of the angular momentum from atomic spins to the body as a whole. 

\section{Acknowledgements}
This work has been supported by the National Science Foundation through grant No. DMR-1161571.

\end{document}